\documentclass[journal]{IEEEtran}

\usepackage[colorlinks]{hyperref}

\usepackage{booktabs}
\usepackage{tabularx}

\usepackage{algorithm}

\usepackage{amssymb}
\usepackage{amsbsy}
\usepackage{amsmath}
\usepackage{bm}
\usepackage{verbatim}
\usepackage{mathrsfs}
\usepackage{amsfonts}
\usepackage{graphicx}
\usepackage[tight,footnotesize]{subfigure}
\usepackage[10pt]{moresize}
\usepackage{array}
\usepackage{color}
\usepackage{epsfig}
\usepackage{stfloats}
\usepackage{balance}

\usepackage{cancel}

\usepackage{hyperref} 

\usepackage{amssymb}

\usepackage[noend]{algpseudocode}
\usepackage{algorithmicx,algorithm}

\usepackage{setspace}
\usepackage{cases}

\usepackage{graphicx}
\usepackage{epstopdf}
\usepackage{multirow}
\usepackage{extarrows}

\usepackage{pifont} 
\usepackage{enumerate}
\usepackage{enumitem}

\newcommand{\subparagraph}{}
\usepackage{titlesec}
\titlespacing{\section}{0pt}{2 ex plus .0ex minus .0ex}{1ex plus .0ex}
\titlespacing{\subsection}{0pt}{1.5 ex plus .0ex minus .0ex}{0.8 ex plus 0.0ex}
\titlespacing{\subsubsection}{0pt}{0.5ex plus .0ex minus .0ex}{0.0ex plus .0ex}

\usepackage{amsmath} 
\allowdisplaybreaks[4]

\ifCLASSINFOpdf

\else

\fi

\begin{document}

\title{Joint Scheduling of Multi-Band Radar Sensing and DNN Inference for Cross-Stage Parallelism}

\author{Yanan~Du,~\IEEEmembership{Member,~IEEE,}
Sai~Xu,~\IEEEmembership{Member,~IEEE,}\\
Kezhi~Wang,~\IEEEmembership{Senior Member,~IEEE,}
and Yansha~Deng,~\IEEEmembership{Senior Member,~IEEE}

\vspace{-3mm}


\thanks{Y.~Du is with the Department of Electronic and Electrical Engineering, University of Sheffield, Sheffield, S1 4ET, U.K. (e-mail: $\rm ynduyndu@163.com$). S. Xu is with University College London, London, U.K. (e-mail: $\rm sai.xu@ucl.ac.uk$). K.~Wang is with the Department of Computer Science, Brunel University
London, UB8 3PH Uxbridge, U.K. (e-mail: $\rm kezhi.wang@brunel.ac.uk$). Y.~Deng is with the Department of Engineering, King’s College
London, WC2R 2LS London, U.K. (e-mail: $\rm yansha.deng@kcl.ac.uk$). (\textit{Corresponding author: Sai~Xu}) }}

\maketitle

\begin{abstract}
This paper studies end-to-end latency minimization for a multi-band radar sensing and deep neural network (DNN) inference pipeline. Unlike conventional stage-wise designs that treat radar sensing and DNN inference as two sequential stages, the proposed framework exploits cross-stage parallelism by allowing the inference branch associated with a sensed band to start as soon as that band completes sensing, without waiting for all bands to finish. To characterize this interaction, we formulate a joint scheduling problem that couples sensing-time allocation, branch release timing, and non-preemptive multi-core execution of a directed acyclic graph (DAG) under sensing-feasibility, precedence, and core-capacity constraints. Since the resulting problem is combinatorial and strongly time-coupled, we further develop a release-aware heuristic that evaluates each sensing decision according to its downstream impact on the DAG makespan, together with a greedy list scheduler for multi-core DAG execution under release times. Simulation results show that the proposed design can effectively exploit cross-stage parallelism and reduce end-to-end latency relative to a decoupled baseline in many heterogeneous sensing scenarios, while also clarifying the operating regimes in which the latency gain becomes limited.
\end{abstract}


\IEEEpeerreviewmaketitle

\section{Introduction}\label{sec:introduction}
\IEEEPARstart{M}{ulti}-band sensing is increasingly integrated with downstream learning-based inference to exploit the complementary information carried by different frequency bands \cite{Smith2024DeepLearning,Liang2024RadarSignal}. Meanwhile, deep neural networks (DNNs) have been shown to effectively fuse multi-band radar or synthetic aperture radar (SAR) observations for improved imaging and classification performance \cite{Smith2024DeepLearning,Liang2024RadarSignal}. In practical spectrum-agile or cognitively managed sensing systems, however, observations from different bands do not necessarily become available simultaneously. Their acquisition may depend on band selection, interference conditions, sensing feasibility, or the need for additional evidence accumulation \cite{Martone2020PracticalAspects,Howard2023HybridCognition}.

Meanwhile, the downstream inference model can often be structured as a multi-branch architecture, in which different bands or modalities are first processed by separate encoders and subsequently fused \cite{Zhao2024DeepMultimodal}. In such an architecture, once the observation from a given band becomes available, the computation associated with the corresponding branch can, in principle, be released before observations from all other bands have been acquired. This perspective is consistent with the broader literature on asynchronous multimodal learning, where heterogeneous inputs may arrive irregularly and update the model state at different times \cite{Gehrig2021CombiningEvents,Shi2024StreamingFlow}.

These observations suggest that end-to-end latency is jointly determined by sensing completion and inference execution, rather than by either stage in isolation. As a result, separately optimizing sensing and computation may lead to poor resource utilization and suboptimal overall performance: a sensing decision that appears locally efficient may delay the release of a branch on the inference critical path, while a compute schedule that is optimal for fixed release times may lose its benefit when the sensing completion order changes. This intuition is consistent with prior work stating that DNN inference can be modeled as a precedence-constrained directed acyclic graph (DAG), whose execution efficiency depends strongly on operator dependencies, task release conditions, and hardware scheduling decisions \cite{You2023AcceleratingConvolutional,Ding2021IOSInterOperator}.

Existing studies have largely treated multi-band sensing and DNN scheduling on multi-core accelerators or heterogeneous hardware as separate problems, while their coupling through branch release times remains underexplored in radar-centric edge inference systems \cite{Smith2024DeepLearning,Gehrig2021CombiningEvents, Gao2026Optimizing,Xu2023CoScheduling}. To address this gap, this paper studies multi-band radar sensing and DNN execution on a multi-core accelerator under a unified timing model. Specifically, sensing follows a time-division policy under time-varying band feasibility, whereas inference is modeled as a precedence-constrained DAG \cite{You2023AcceleratingConvolutional,Ding2021IOSInterOperator}. In this formulation, the completion time of each sensing band serves as the release time of its corresponding inference branch. The objective is to minimize the end-to-end latency from sensing initiation to final inference output through joint coordination of band activation and DAG execution.

\subsection{Related Work}\label{sec:relatedwork}
Prior work lies in two largely separate research threads: multi-band radar sensing and accelerator-level DNN scheduling. Our work is motivated by the observation that the missing link between them is the release-time coupling induced by progressive band completion.

\subsubsection{Multi-Band Radar Sensing}
Multi-band radar has been widely studied due to its ability to exploit the complementary properties of different frequency bands, including penetration capability, spatial resolution, scattering behavior, and target representation. Existing work has leveraged this advantage in various applications, such as high-resolution 3-D SAR imaging for concealed and occluded object detection \cite{Smith2024DeepLearning}, multi-band SAR/inverse synthetic aperture radar (ISAR) for high-resolution imaging and target detection through coherent fusion \cite{Zhong2024MultipleInOne}, weather radar data fusion and super-resolution for severe weather monitoring \cite{Yang2026MultiChannelSuperResolution}, and dual-band SAR imagery for foliage-penetrating target detection \cite{Zhang2024DualBandSAR}. With the advancement of deep learning, learning-based methods have been increasingly introduced into multi-band radar processing to capture cross-band relationships and improve information fusion. Representative applications include multi-subband radar signal fusion \cite{Jiang2022MultiSubbandRadar}, small-target detection in imaging radar \cite{Gong2024MultibandRadar}, multi-band SAR image classification \cite{Zhu2022SARImage}, and multi-band polarimetric synthetic aperture radar (PolSAR) land-cover classification \cite{Han2025ViTKAN}. In particular, some recent studies have adopted multi-branch DNNs, in which each band is first processed by a dedicated branch before cross-band fusion. This strategy has been used in dual-band high-resolution range profile (HRRP) recognition \cite{Yang2023DualBandPolarimetric,Yang2024DualBandHRRP,Wang2024DualBandHRRP} and dual-frequency HRRP recognition with missing-modality completion \cite{Zhou2024MissingModality}. However, these studies mainly emphasize sensing quality and inference accuracy, rather than end-to-end scheduling under asynchronous band completion.

\subsubsection{DNN Execution on Accelerators}
Efficient execution of DNN workloads on multi-core accelerators has been widely studied in the computer architecture community, driven by the need to improve throughput and resource utilization under increasingly heterogeneous workloads. Existing work has addressed this problem through hand-crafted layer-level scheduling for multi-DNN inference \cite{Baek2020MultiNeural}, genetic-algorithm-based schedule exploration with memory-aware optimization \cite{Kao2022MagmaOptimization,Zheng2023MemoryComputation}, and dynamic memory management for concurrent DNN execution \cite{Kim2023MoCAMemoryCentric}. For edge scenarios, prior studies have further investigated real-time multi-DNN execution through heterogeneous dataflow designs and adaptive online scheduling \cite{Kwon2021HeterogeneousDataflow,Kim2023DREAMDynamic}, while sparse multi-DNN workloads and learning-based scheduling have also been considered in DySta \cite{Fan2023SparseDySta} and TaiChi \cite{Zhou2025TaiChiEfficient}, respectively. These studies clearly demonstrate the importance of precedence-aware execution and memory locality, but they typically assume that all input tensors required by the workload are already available. As a result, they do not model the release-time dependencies created by progressive radar sensing completion.

In summary, prior work has not explicitly addressed the following question: when different radar bands complete sensing at different times, how should one jointly schedule band activation and DAG execution so as to minimize end-to-end latency? This paper addresses exactly this release-aware cross-stage scheduling problem.

\subsection{Contributions}
The serial, stage-wise execution of sensing and DNN execution limits cross-stage parallelism, causing some computations that could otherwise start earlier to be delayed and resulting in low utilization of computing resources. Motivated by this observation, this paper proposes a cross-stage parallel processing approach for multi-band radar sensing and DNN execution to reduce end-to-end latency. The main contributions of this work are summarized as follows.
\begin{itemize}[leftmargin=*]
\item We develop a unified timing model for multi-band radar sensing and DNN inference in which the sensing completion time of each band explicitly determines the release time of the corresponding DNN branch-entry job. This model removes the ambiguity between sensing and computation stages and directly captures the cross-stage coupling of interest.
\item We formulate an end-to-end latency minimization problem that jointly optimizes radar-band activation and multi-core DAG execution under sensing-feasibility, precedence, and non-overlap constraints. We further explain how the original mixed discrete problem is approximated by a release-aware rollout heuristic, thereby bridging the gap between the optimization model and the practical scheduler.
\item We propose a release-aware heuristic consisting of two components: a greedy DAG scheduler for non-preemptive multi-core execution under release times, and a sensing-side one-step look-ahead rule that evaluates each candidate band by its downstream impact on the DAG makespan.
\item We conduct comparative simulations against a decoupled baseline under diverse system settings, including sensing-information thresholds, branch-node distributions, bandwidth, signal-to-interference-plus-noise ratio (SINR) thresholds, and accelerator core counts. The results show that the proposed release-aware design more effectively exploits cross-stage parallelism and clarify the conditions under which it yields significant latency gains, limited improvement, or even increased latency.
\end{itemize}
\par The rest of this paper is organized as follows. Section~\ref{sec:system_model} presents the system model and the joint latency-minimization problem. Section~\ref{sec:proposed_method} introduces the proposed scheduling method and analyzes its complexity. Section~\ref{sec:simulation_results} reports the simulation results and sensitivity analysis under different operating configurations. Section~\ref{sec:conclusions} concludes this paper.

\section{System Model}\label{sec:system_model}

\begin{figure}
\centering
\includegraphics[width=\linewidth]{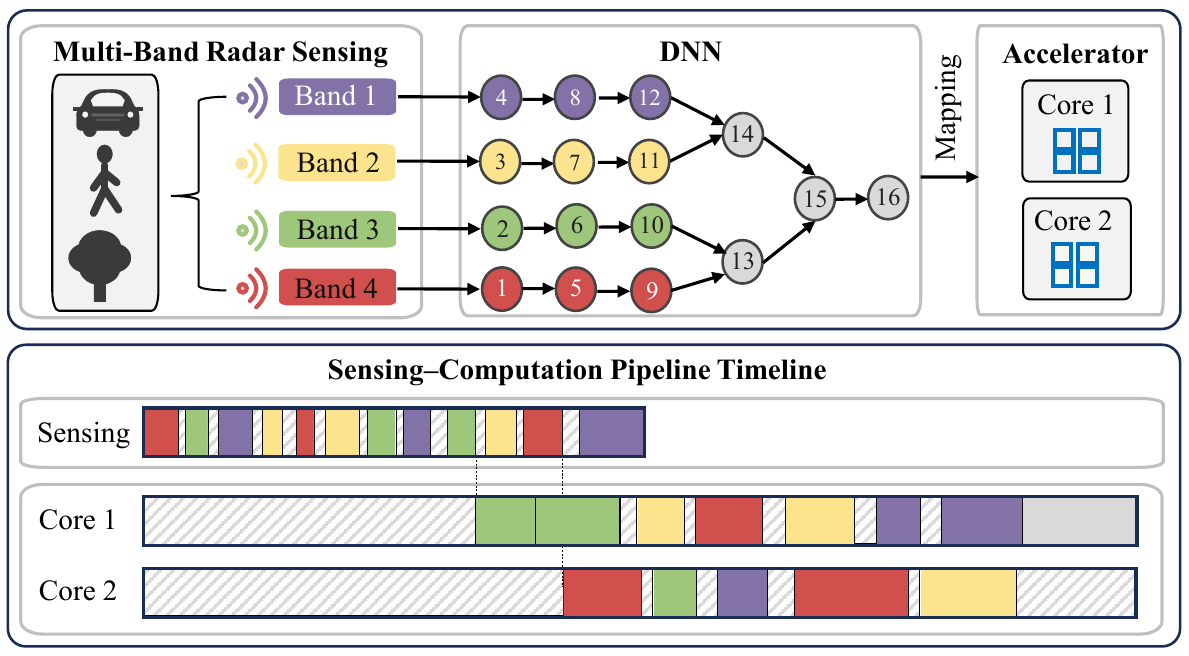}
\caption{System architecture of latency-aware joint scheduling for multi-band radar sensing and DNN inference.}\label{Fig1}
\end{figure}

Fig.~\ref{Fig1} illustrates a latency-aware joint scheduling framework for multi-band radar sensing and DNN inference. The system coordinates multi-band radar front-end sensing and DNN inference under latency constraints. Specifically, heterogeneous observations are acquired across different radar bands according to a joint scheduling policy, and then delivered to a DNN for cross-band fusion and task-oriented inference, aiming at reducing end-to-end latency while maintaining inference performance.
\subsection{Multi-Band Radar Sensing Model}

Consider a radar system operating over multiple frequency bands, indexed by \(k \in \mathcal{K} \triangleq \{1,2,\dots,K\}\). Each band occupies distinct spectral resource and is capable of independently acquiring sensing information about the environment or targets. To avoid simultaneous use of
multiple bands, a time-division sensing mechanism is adopted, under which at most one band can be activated in each discrete time slot. Let \(a_{k,t}\in\{0,1\}\) denote the activation indicator of band \(k\) at slot \(t\), where \(a_{k,t}=1\) means that band \(k\) is selected for sensing and \(a_{k,t}=0\) otherwise. The time-division constraint is therefore given by
\begin{equation}
\sum_{k=1}^{K} a_{k,t} \le 1, \qquad \forall t .
\label{eq:tdm_constraint}
\end{equation}

Due to time-varying propagation and environmental conditions, the sensing quality of each band may change over time. Let \(\gamma_k(t)\) denote the instantaneous sensing SINR of band \(k\) at time \(t\), and let \(\Gamma_k(t)\) denote the corresponding minimum SINR threshold required to satisfy
a prescribed detection-performance criterion. A band is regarded as feasible only if its current sensing quality exceeds the required threshold, i.e.,
\begin{equation}
\gamma_k(t) > \Gamma_k(t).
\end{equation}
Accordingly, the feasible band set at time \(t\) is defined as
\begin{equation}
\mathcal{A}(t)=\left\{\, k\in\mathcal{K}\;\middle|\; \gamma_k(t)>\Gamma_k(t) \,\right\}.
\label{eq:feasible_set}
\end{equation}
Only feasible bands are eligible for activation. Hence, the sensing decision must satisfy
\begin{equation}
a_{k,t} \le \mathbf{1}\{k\in\mathcal{A}(t)\}, \qquad \forall k\in\mathcal{K},\ \forall t,
\label{eq:feasible_activation}
\end{equation}
where \(\mathbf{1}\{\cdot\}\) denotes the indicator function. When \(\mathcal{A}(t)=\varnothing\), no band can be selected and the radar remains idle in slot \(t\).

When a band is activated, it contributes a certain amount of useful sensing information in that slot. Let \(X_k(t)\ge 0\) denote the accumulated sensing information collected from band \(k\) up to slot \(t\), and let \(\eta_k>0\) denote the target information threshold required for completing sensing on
band \(k\). The information accumulation process is modeled as
\begin{equation}
X_k(t+1)=X_k(t)+a_{k,t}\,g_k\!\left(\gamma_k(t)\right),
\label{eq:info_update}
\end{equation}
where \(g_k(\cdot)\) is a nondecreasing function that characterizes the effective sensing gain of band \(k\) as a function of the instantaneous quality metric. A commonly adopted form is
\begin{equation}
g_k\!\left(\gamma_k(t)\right)= B_k \log\! \bigl(1+\gamma_k(t)\bigr),
\label{eq:gain_function}
\end{equation}
where $B_k$ denotes the bandwidth of band \(k\). Band \(k\) is considered to have completed its sensing task once the accumulated information reaches the required threshold, namely,
\begin{equation}
X_k(t)\ge \eta_k .
\label{eq:info_completion}
\end{equation}
Additionally, once the accumulated sensing information of band \(k\) reaches \(\eta_k\), the corresponding radar observation is regarded as available and the associated branch-entry job in the DNN becomes releasable.
\subsection{DNN Inference Execution Model}
In the considered system, DNN inference can be launched before sensing over all radar bands is fully completed. Specifically, the DNN can be represented by a DAG, denoted by \(G=(\mathcal{V},\mathcal{E})\). Each vertex \(v\in\mathcal{V}\) corresponds to a computational task, while each directed edge \((u,v)\in\mathcal{E}\) indicates that task \(v\) depends on the output of task \(u\). Hence, a task becomes executable only after all of its predecessor tasks have completed. To minimize the end-to-end latency from sensing to inference, the inference pipeline may be initiated once sufficient sensing data from any one band become available as a data slice, instead of waiting for the full set of multi-band observations. This allows sensing and inference to proceed in parallel, thereby lowering the overall end-to-end delay.

The DNN is executed on a multi-core hardware accelerator, where each core consists of an array of processing elements (PEs). For layer-level scheduling, we introduce the concepts of \emph{job} and \emph{mapping}. A job is defined as the minimum execution unit that can be independently dispatched to the accelerator, whereas a mapping specifies both the core assignment of jobs and their execution order over time. Let \(\mathcal{C}=\{1,\ldots,C\}\) denote the set of accelerator cores. For each job \(v\in\mathcal{V}\), define a binary variable \(y_{v,c}\in\{0,1\}\), where \(y_{v,c}=1\) indicates that job \(v\) is executed on core \(c\in\mathcal{C}\). Each job is assigned to exactly one core, i.e.,
\begin{equation}
\sum_{c\in\mathcal{C}} y_{v,c} = 1, \qquad \forall v\in\mathcal{V}.
\label{eq:core_assignment}
\end{equation}

Let \(p_{v,c}\) denote the effective execution time of job \(v\) on core \(c\). This latency depends on both computation and memory access, and is determined by the actual execution mapping and data residency status in the memory hierarchy. In particular, different memory-access overheads may arise depending on whether the data required by job \(v\) are served from on-chip buffers or fetched from off-chip memory. Therefore, \(p_{v,c}\) is treated as a mapping-dependent execution cost that captures both arithmetic processing and the corresponding data-movement overhead. Let \(s_v\) and \(f_v\) represent its start time and completion time, respectively. Since job execution is assumed to be non-preemptive, the completion time of job \(v\) can be written as
\begin{equation}
f_v = s_v + \sum_{c\in\mathcal{C}} y_{v,c}\, p_{v,c}, \qquad \forall v\in\mathcal{V}.
\label{eq:finish_time}
\end{equation}

The DAG structure imposes precedence constraints on the execution order. In particular, a job cannot start until all of its immediate predecessors have finished. Therefore,
\begin{equation}
s_v \ge f_u, \qquad \forall (u,v)\in\mathcal{E}.
\label{eq:precedence}
\end{equation}
In addition to dependency constraints, the hardware architecture also limits the degree of parallelism. Because each accelerator core can process at most one job at any time, the set of jobs assigned to the same core must form a feasible non-overlapping schedule. To avoid introducing pairwise ordering variables and big-\(M\) constraints, we express this requirement in a compact feasibility form as
\begin{equation}
\left\{(s_v,f_v)\mid y_{v,c}=1,\ v\in\mathcal{V}\right\} \in \mathcal{F}_c,
\qquad \forall c\in\mathcal{C},
\label{eq:core_feasibility}
\end{equation}
where \(\mathcal{F}_c\) denotes the set of all non-preemptive schedules on core \(c\) with no temporal overlap among assigned jobs. Equivalently, for any two distinct jobs \(u\neq v\) satisfying \(y_{u,c}=y_{v,c}=1\), one must have either \(f_u \le s_v\) or \(f_v \le s_u\).

\subsection{Optimization Problem}
The objective is to minimize the overall end-to-end latency, defined as the completion time of the entire inference process. Since the DNN output becomes available only after all scheduled jobs have finished execution, the total latency can be expressed as
\begin{equation}
T_{\mathrm{total}}=\max_{v\in\mathcal{V}} ~f_v .
\label{eq:total_latency}
\end{equation}
The end-to-end latency depends on both the sensing schedule and the execution schedule of DNN jobs. On the sensing side, the band activation variables \(\{a_{k,t}\}\) determine when sufficient information is accumulated for each band. On the computation side, the mapping variables \(\{y_{v,c}\}\) together with the start and finish times \(\{s_v,f_v\}\) determine how the DNN workload is executed on the multi-core accelerator. Therefore, the joint scheduling problem is formulated as
\begin{equation}
\begin{aligned}
\min_{\{a_{k,t}\},\,\{y_{v,c}\},\,\{s_v,f_v\}} \quad
& T_{\mathrm{total}}\\
\text{s.t.}\quad
& \eqref{eq:tdm_constraint},\ \eqref{eq:feasible_activation},\ \eqref{eq:info_completion},  \\
& \eqref{eq:core_assignment},\ \eqref{eq:precedence},\ \eqref{eq:core_feasibility}.
\end{aligned}
\label{eq:joint_problem}
\end{equation}
\section{Proposed Algorithm}\label{sec:proposed_method}
Since the sensing decisions and the DNN execution mapping are tightly coupled over time, solving \eqref{eq:joint_problem} optimally is challenging. In particular, the completion time of sensing on each band determines when the corresponding branch of the inference DAG can be released, while the subsequent execution latency depends on core assignment, execution order, and memory-access overhead. The resulting problem is a mixed discrete optimization with temporal coupling, which is intractable for large-scale instances. Another major challenge stems from the non-anticipative online nature of the problem: at each slot, only the current SINR is available, while future SINR values remain unknown. As a result, slot-wise decisions must be made causally, even though the final latency depends on the entire sequence of sensing and execution decisions across slots. To enable efficient online scheduling, we develop a greedy joint scheduling algorithm that coordinates band activation and DAG execution in a rollout manner. As a benchmark, we also consider a decoupled scheme in which sensing and inference are scheduled independently.

The role of the proposed algorithm is therefore to compute a feasible heuristic solution to \eqref{eq:joint_problem}. More specifically, the original problem jointly optimizes sensing variables \(\{a_{k,t}\}\), core-assignment variables \(\{y_{v,c}\}\), and job timing variables \(\{s_v,f_v\}\), which are tightly coupled through branch release times. Solving this mixed discrete problem exactly is intractable even for moderate graph sizes. We therefore retain the most critical coupling---namely, the dependence of DAG makespan on sensing-induced branch release times---and approximate the original problem through a two-level heuristic: (i) given release times, a greedy list scheduler constructs a feasible DAG schedule on the accelerator; and (ii) at each sensing slot, a one-step rollout rule evaluates candidate band activations using the downstream makespan returned by the scheduler. In this sense, the proposed method is a release-aware approximation to \eqref{eq:joint_problem}, rather than an exact optimizer.

\subsection{Release-Time Modeling for Multi-Band DAG Inference}
To explicitly couple sensing completion with DNN execution, we associate each sensing band \(k\in\mathcal{K}\) with an entry job \(r_k \in \mathcal{V}\) in the DAG, corresponding to the first executable node of the inference branch triggered by band \(k\). Let
\begin{equation}
\tau_k \triangleq \min \left\{ t ~\middle|~ X_k(t)\ge \eta_k \right\}
\label{eq:band_completion_time}
\end{equation}
denote the sensing completion time of band \(k\). Then, the entry job \(r_k\) cannot start before \(\tau_k\), i.e.,
\begin{equation}
s_{r_k} \ge \tau_k, \qquad \forall k\in\mathcal{K}.
\label{eq:release_constraint}
\end{equation}
For all non-entry jobs, the release condition is implicitly captured by DAG precedence constraints. Therefore, the sensing process affects inference execution only through the release times of the entry jobs. This release-time abstraction matches the progressive execution behavior considered in the system model: once sensing on one band is completed, the corresponding branch of the DNN can start immediately, even if other bands are still being used for sensing. In this way, sensing and inference can overlap in time.

\subsection{Mapping-Dependent Job Latency Model}

The latency associated with a job consists of its computation time and the mapping-dependent delay required to make predecessor data available. To capture this effect, we distinguish processor occupancy from inter-job data-access latency.

For each job \(v\in\mathcal{V}\), let \(d_v^{\mathrm{cmp}}\) denote its pure computation time. In addition, let \(r_v^{\mathrm{on}}\) and \(r_v^{\mathrm{off}}\) denote the latency for reading the input of job \(v\) from on-chip and off-chip memory, respectively. Similarly, let
\(w_u^{\mathrm{on}}\) and \(w_u^{\mathrm{off}}\) denote the latency for writing the output of job \(u\) to on-chip and off-chip storage, respectively. We assume that on-chip forwarding is possible only when two dependent jobs are mapped to the same core and remain in the same branch; otherwise, the output of \(u\) must be spilled to off-chip memory and later fetched by \(v\). Define
\begin{equation}
\delta_{uv}=
\begin{cases}
1, & m(u)=m(v),\ b(u)=b(v),\\
0, & \text{otherwise},
\end{cases}
\end{equation}
where \(m(v)\in\mathcal{C}\) denotes the core assigned to job \(v\), and \(b(v)\) its branch index. Then, for each dependency edge \((u,v)\in\mathcal{E}\), the data-ready time of \(u\)'s output for job \(v\) is
\begin{equation}
\phi_{u\rightarrow v}
=
f_u
+\delta_{uv}\!\left(w_u^{\mathrm{on}}+r_v^{\mathrm{on}}\right)
+\left(1-\delta_{uv}\right)\!\left(w_u^{\mathrm{off}}+r_v^{\mathrm{off}}\right).
\end{equation}
Accordingly, the start time of job \(v\) must satisfy
\begin{equation}
s_v \ge \max\!\left\{ \rho_v,\ \chi_{m(v)},\ \max_{u\in\mathrm{pred}(v)}
\phi_{u\rightarrow v} \right\},
\end{equation}
where \(\rho_v\) is the release time of job \(v\), and \(\chi_{m(v)}\) is the time at which core \(m(v)\) becomes idle. The completion time of \(v\) is then
\begin{equation}
f_v = s_v + d_v^{\mathrm{cmp}}.
\end{equation}

This formulation makes the effective latency explicitly mapping-dependent: co-locating dependent jobs can reduce data-access delay, whereas separating them across cores may expose more parallelism by reducing contention for processor time. The scheduler must therefore balance locality against concurrency.

\subsection{Greedy DAG Scheduling with Release Times}

Given the sensing completion times \(\{\tau_k\}\), we consider the inference-side subproblem of scheduling the DAG on \(C\) accelerator cores. Since the graph is acyclic, jobs can be processed in a topological order. For each job \(v\), the scheduler evaluates all candidate cores and selects the one yielding the earliest completion time.

Let \(\pi=(v_1,\dots,v_{|\mathcal{V}|})\) denote a topological ordering of \(G\). For each job \(v_i\) and each core \(c\in\mathcal{C}\), we compute a candidate start time
\begin{equation}
\hat{s}_{v_i,c}
=
\max\!\left\{
\chi_c,\,
\rho_{v_i},\,
\max_{u\in\mathrm{pred}(v_i)} \hat{\phi}_{u\rightarrow v_i}^{(c)}
\right\},
\label{eq:candidate_start}
\end{equation}
where \(\hat{\phi}_{u\rightarrow v_i}^{(c)}\) is the predecessor data-ready time assuming that \(v_i\) is placed on core \(c\). The corresponding candidate finish time is
\begin{equation}
\hat{f}_{v_i,c} = \hat{s}_{v_i,c} + d_{v_i}^{\mathrm{cmp}}.
\label{eq:candidate_finish}
\end{equation}
The selected core is then
\begin{equation}
m(v_i)=\arg\min_{c\in\mathcal{C}}~\hat{f}_{v_i,c},
\label{eq:core_selection_rule}
\end{equation}
and the job timing is updated as
\begin{equation}
s_{v_i}=\hat{s}_{v_i,m(v_i)}, \qquad
f_{v_i}=\hat{f}_{v_i,m(v_i)}.
\label{eq:selected_timing}
\end{equation}
After scheduling \(v_i\), the availability time of core \(m(v_i)\) is set to \(f_{v_i}\).

This procedure yields a feasible non-preemptive schedule satisfying both the precedence and core-capacity constraints. Its complexity is polynomial in the graph size and the number of cores, making it suitable for repeated use inside the joint scheduler.

\subsection{Joint Greedy Sensing-and-Execution Scheduling}
The key idea of joint greedy sensing-and-execution scheduling is that each sensing decision is evaluated not only by its immediate sensing gain, but also by its downstream impact on the final DAG completion time.

At each slot \(t\), the feasible band set \(\mathcal{A}(t)\) is first determined according to \eqref{eq:feasible_set}. If \(\mathcal{A}(t)=\varnothing\), the radar remains idle in that slot. Otherwise, for each feasible band \(k\in\mathcal{A}(t)\) that has not yet completed sensing, we tentatively assume that the current slot is allocated to band \(k\). Let \(\Delta X_k(t)=g_k(\gamma_k(t))\) denote the sensing information obtained if band \(k\) is activated at slot \(t\). The temporary sensing state becomes
\begin{equation}
\tilde{X}_j(t+1)=
\begin{cases}
X_k(t)+\Delta X_k(t), & j=k,\\
X_j(t), & j\neq k.
\end{cases}
\label{eq:tentative_info_update}
\end{equation}
If this tentative update makes \(\tilde{X}_k(t+1)\ge \eta_k\), the release time of the corresponding entry job is set to the current slot:
\begin{equation}
\tilde{\tau}_k=t.
\label{eq:tentative_release}
\end{equation}

All other release times remain unchanged. Based on the resulting temporary release vector \(\tilde{\bm{\tau}}\), the greedy DAG scheduler described above is invoked to estimate the resulting total latency,
\begin{equation}
\tilde{T}_{\mathrm{total}}^{(k)} = \mathcal{S}\!\left(G,\tilde{\bm{\tau}}\right),
\label{eq:rollout_estimate}
\end{equation}
where \(\mathcal{S}(\cdot)\) denotes the makespan returned by the greedy DAG scheduling routine. 
The current slot is then assigned to the band that yields the smallest estimated end-to-end latency:
\begin{equation}
k_t^\star
=
\arg\min_{k\in\mathcal{A}(t),\, X_k(t)<\eta_k}
\tilde{T}_{\mathrm{total}}^{(k)}.
\label{eq:joint_decision_rule}
\end{equation}
After the decision is made, the sensing state is updated according to the selected band, and the process proceeds to the next slot. Once all bands satisfy \eqref{eq:info_completion}, the final release times \(\{\tau_k\}\) are obtained, and the complete execution schedule is generated by the greedy DAG scheduler.

The above procedure is a one-step look-ahead policy. Although it does not solve the joint problem globally, it explicitly accounts for the interaction between sensing completion and DAG execution at every slot, and therefore captures the most critical coupling neglected by conventional decoupled methods.

\subsection{Independent Greedy Baseline}
To benchmark the benefit of joint optimization, we also consider a decoupled baseline in which sensing scheduling and DNN execution scheduling are performed independently.
On the sensing side, each slot is allocated to the feasible unfinished band with the largest remaining sensing demand. Specifically, define the residual sensing requirement of band \(k\) at slot \(t\) as
\begin{equation}
R_k(t)=\max\{0,\eta_k-X_k(t)\}.
\label{eq:residual_demand}
\end{equation}
Among all feasible bands in \(\mathcal{A}(t)\), the selected band is
\begin{equation}
k_t^{\mathrm{ind}}
=
\arg\max_{k\in\mathcal{A}(t)} R_k(t).
\label{eq:independent_sensing_rule}
\end{equation}
This rule is purely sensing-driven and does not consider the downstream effect on inference latency. After the sensing completion times \(\{\tau_k\}\) are determined, the same greedy DAG scheduler is applied once to obtain the final execution mapping and total latency.
Hence, the difference between the proposed method and the baseline lies entirely in the sensing-stage decision rule: the former minimizes a rollout estimate of the overall end-to-end latency, whereas the latter greedily reduces sensing backlog without considering the inference DAG.

\algrenewcommand\algorithmicrequire{\textbf{Input:}}
\algrenewcommand\algorithmicensure{\textbf{Output:}}

\begin{algorithm}[t]
\caption{Release-Aware Greedy DAG (RADG)}
\label{alg:dag_greedy_release}
\begin{algorithmic}[1]
\Require \(G=(\mathcal{V},\mathcal{E})\); \(\{\rho_v\}_{v\in\mathcal{V}}\); \(C\)
\Ensure \(\{m(v)\}_{v\in\mathcal{V}}\); \(\{s_v\}_{v\in\mathcal{V}}\); \(\{f_v\}_{v\in\mathcal{V}}\); \(L\)
\State Obtain a topological order \(\Pi=(v_1,\dots,v_{|\mathcal{V}|})\) of \(G\)
\State Initialize core-availability times \(\chi_c \gets 0,\ \forall c\in\mathcal{C}\)
\ForAll{\(v\in\Pi\)}
    \State \(c^\star \gets 1,\ \hat{f}^\star \gets +\infty,\ \hat{s}^\star \gets 0\)
    \For{each core \(c\in\mathcal{C}\)}
        \State Compute \(\hat{\phi}^{(c)}_{u\rightarrow v}\) for all \(u\in\mathrm{pred}(v)\)
        \State \(\hat{s}_{v,c}\gets \max\!\left\{\chi_c,\rho_v,\max_{u\in\mathrm{pred}(v)}\hat{\phi}^{(c)}_{u\rightarrow v}\right\}\)
        \State \(\hat{f}_{v,c}\gets \hat{s}_{v,c}+d_v^{\mathrm{cmp}}\)
        \If{\(\hat{f}_{v,c}<\hat{f}^\star\)}
            \State \(\hat{f}^\star\gets \hat{f}_{v,c},\ \hat{s}^\star\gets \hat{s}_{v,c},\ c^\star\gets c\)
        \EndIf
    \EndFor
    \State \(m(v)\gets c^\star,\ s_v\gets \hat{s}^\star,\ f_v\gets \hat{f}^\star\)
    \State \(\chi_{c^\star}\gets f_v\)
\EndFor
\State \(L\gets \max_{v\in\mathcal{V}} f_v\)
\State \Return \(\{m(v)\}_{v\in\mathcal{V}},\{s_v\}_{v\in\mathcal{V}},\{f_v\}_{v\in\mathcal{V}},L\)
\end{algorithmic}
\end{algorithm}

\begin{algorithm}[t]
\caption{Joint Scheduling}
\label{alg:joint_greedy}
\begin{algorithmic}[1]
\Require \(\mathcal{K}\); \(T_{\max}\); \(\{\eta_k\}_{k\in\mathcal{K}}\); \(\{g_k(\cdot)\}_{k\in\mathcal{K}}\); \(\{\mathcal{A}(t)\}_{t=1}^{T_{\max}}\); \(G=(\mathcal{V},\mathcal{E})\); \(\{v_k^{\mathrm{ent}}\}_{k\in\mathcal{K}}\); \(C\)
\Ensure \(\{a_{k,t}\}\); \(\{\tau_k\}\); \(\{m(v),s_v,f_v\}\); \(L\)
\State Initialize \(X_k(1)\gets 0,\ \tau_k\gets +\infty,\ \forall k\in\mathcal{K}\)
\State Initialize \(a_{k,t}\gets 0,\ \forall k\in\mathcal{K},\ t=1,\dots,T_{\max}\)
\For{\(t=1\) to \(T_{\max}\)}
    \If{\(X_k(t)\ge \eta_k,\ \forall k\in\mathcal{K}\)}
        \State \textbf{break}
    \EndIf
    \State \(\mathcal{A}_{\mathrm{u}}(t)\gets \{k\in\mathcal{A}(t)\mid X_k(t)<\eta_k\}\)
    \If{\(\mathcal{A}_{\mathrm{u}}(t)=\varnothing\)}
        \State continue
    \EndIf
    \State \(k^\star\gets \bot,\ \tilde{L}^\star\gets +\infty\)
    \ForAll{\(k\in\mathcal{A}_{\mathrm{u}}(t)\)}
        \State Tentatively set \(\tilde{X}_j\gets X_j(t)\) and \(\tilde{\tau}_j\gets \tau_j,\ \forall j\in\mathcal{K}\)
        \State \(\tilde{X}_k\gets X_k(t)+g_k(\gamma_k(t))\)
        \If{\(\tilde{X}_k\ge \eta_k\) and \(\tilde{\tau}_k=+\infty\)}
            \State \(\tilde{\tau}_k\gets t\)
        \EndIf
        \State Set \(\rho_v\gets 0,\ \forall v\in\mathcal{V}\)
        \ForAll{\(j\in\mathcal{K}\)}
            \If{\(\tilde{\tau}_j<+\infty\)}
                \State \(\rho_{v_j^{\mathrm{ent}}}\gets \tilde{\tau}_j\)
            \Else
                \State \(\rho_{v_j^{\mathrm{ent}}}\gets +\infty\)
            \EndIf
        \EndFor
        \State \((\{m(v)\},\{s_v\},\{f_v\},\tilde{L})\gets \textsc{RADG}(G,\{\rho_v\},C)\)
        \If{\(\tilde{L}<\tilde{L}^\star\)}
            \State \(\tilde{L}^\star\gets \tilde{L},\ k^\star\gets k\)
        \EndIf
    \EndFor
    \If{\(k^\star\neq \bot\)}
        \State \(a_{k^\star,t}\gets 1\)
        \State \(X_{k^\star}(t+1)\gets X_{k^\star}(t)+g_{k^\star}(\gamma_{k^\star}(t))\)
        \ForAll{\(j\in\mathcal{K}\setminus\{k^\star\}\)}
            \State \(X_j(t+1)\gets X_j(t)\)
        \EndFor
        \If{\(X_{k^\star}(t+1)\ge \eta_{k^\star}\) and \(\tau_{k^\star}=+\infty\)}
            \State \(\tau_{k^\star}\gets t\)
        \EndIf
    \EndIf
\EndFor
\State Set final release times \(\rho_v\gets 0,\ \forall v\in\mathcal{V}\), and \(\rho_{v_k^{\mathrm{ent}}}\gets \tau_k,\ \forall k\in\mathcal{K}\)
\State \((\{m(v)\},\{s_v\},\{f_v\},L)\gets \textsc{RADG}(G,\{\rho_v\},C)\)
\State \Return \(\{a_{k,t}\},\{\tau_k\},\{m(v),s_v,f_v\},L\)
\end{algorithmic}
\end{algorithm}

\begin{algorithm}[t]
\caption{Decoupled Scheduling}
\label{alg:independent_greedy}
\begin{algorithmic}[1]
\Require \(\mathcal{K}\); \(T_{\max}\); \(\{\eta_k\}_{k\in\mathcal{K}}\); \(\{g_k(\cdot)\}_{k\in\mathcal{K}}\); \(\{\mathcal{A}(t)\}_{t=1}^{T_{\max}}\); \(G=(\mathcal{V},\mathcal{E})\); \(\{v_k^{\mathrm{ent}}\}_{k\in\mathcal{K}}\); \(C\)
\Ensure \(\{a_{k,t}\}\); \(\{\tau_k\}\); \(\{m(v),s_v,f_v\}\); \(L\)
\State Initialize \(X_k(1)\gets 0,\ \tau_k\gets +\infty,\ \forall k\in\mathcal{K}\)
\State Initialize \(a_{k,t}\gets 0,\ \forall k\in\mathcal{K},\ t=1,\dots,T_{\max}\)
\For{\(t=1\) to \(T_{\max}\)}
    \If{\(X_k(t)\ge \eta_k,\ \forall k\in\mathcal{K}\)}
        \State \textbf{break}
    \EndIf
    \State \(\mathcal{A}_{\mathrm{u}}(t)\gets \{k\in\mathcal{A}(t)\mid X_k(t)<\eta_k\}\)
    \If{\(\mathcal{A}_{\mathrm{u}}(t)=\varnothing\)}
        \State \textbf{continue}
    \EndIf
    \State \(k^\star\gets \arg\max_{k\in\mathcal{A}_{\mathrm{u}}(t)} \left(\eta_k-X_k(t)\right)\)
    \State \(a_{k^\star,t}\gets 1\)
    \State \(X_{k^\star}(t+1)\gets X_{k^\star}(t)+g_{k^\star}(\gamma_{k^\star}(t))\)
    \ForAll{\(j\in\mathcal{K}\setminus\{k^\star\}\)}
        \State \(X_j(t+1)\gets X_j(t)\)
    \EndFor
    \If{\(X_{k^\star}(t+1)\ge \eta_{k^\star}\) and \(\tau_{k^\star}=+\infty\)}
        \State \(\tau_{k^\star}\gets t\)
    \EndIf
\EndFor
\State \(t_{\mathrm{start}}\gets \max_{k\in\mathcal{K}}\tau_k\)
\State Set \(\rho_v\gets 0,\ \forall v\in\mathcal{V}\), and \(\rho_{v_k^{\mathrm{ent}}}\gets t_{\mathrm{start}},\ \forall k\in\mathcal{K}\)
\State \((\{m(v)\},\{s_v\},\{f_v\},L)\gets \textsc{RADG}(G,\{\rho_v\},C)\)
\State \Return \(\{a_{k,t}\},\{\tau_k\},\{m(v),s_v,f_v\},L\)
\end{algorithmic}
\end{algorithm}

\subsection{Discussion on Practical Implementation}

The proposed framework supports progressive execution in practical multi-band sensing systems. Once any band finishes sensing, the corresponding inference branch can be released immediately on the accelerator, allowing the system to overlap front-end sensing and back-end DNN processing. This is particularly beneficial when the DAG contains branch-specific feature extractors followed by cross-band alignment and fusion modules, as early branch execution can hide part of the sensing latency of other bands.
Moreover, the mapping-dependent latency model naturally captures an important architectural tradeoff. Assigning dependent jobs to the same core improves on-chip data reuse and reduces memory traffic, while distributing jobs across different cores enables parallel execution but may incur off-chip transfers. The proposed greedy core-selection rule explicitly balances these two effects at runtime.
Finally, the rollout-based joint scheduler is computationally light compared with exhaustive search. At each slot, it evaluates only the currently feasible candidate bands and invokes a polynomial-time DAG scheduler for each candidate. Therefore, it provides a practical compromise between performance and complexity for latency-sensitive radar sensing-and-computing systems.

\subsection{Proposed Scheduling Algorithms and Complexity Analysis}
Algorithm \ref{alg:dag_greedy_release} presents the release-aware greedy DAG scheduler used in the computation stage, where each node is assigned to the core that yields the earliest finish time under precedence constraints, core availability, and node release times. Based on this routine, Algorithm \ref{alg:joint_greedy} develops a joint greedy scheduling scheme that determines each sensing decision by minimizing the estimated end-to-end latency with sensing and computation considered together. In contrast, Algorithm \ref{alg:independent_greedy}  adopts a decoupled design, where sensing resources are greedily allocated according to the remaining sensing demand, and the DAG is scheduled only after all slices have completed sensing via Algorithm \ref{alg:dag_greedy_release}.

Let \(K=|\mathcal{K}|\), \(V=|\mathcal{V}|\), and \(C=|\mathcal{C}|\) denote the numbers of sensing bands, DAG jobs, and accelerator cores, respectively. In the greedy DAG scheduler, each job is evaluated on all \(C\) cores, and for each candidate core the scheduler checks all predecessor jobs. Denoting by \(E=|\mathcal{E}|\) the number of DAG edges, the complexity of one DAG scheduling pass is \(O(C(V+E))\). For the proposed joint scheduler, each slot evaluates at most \(K\) feasible candidate bands, and each evaluation requires one invocation of the greedy DAG scheduler. If the sensing process spans \(T\) slots, the overall complexity is $O\!\left(TK C(V+E)\right)$.
In contrast, the decoupled baseline performs only one sensing decision per slot and invokes the DAG scheduler once after sensing is completed, yielding complexity $O(TK + C(V+E))$. 
\section{Simulation Results}\label{sec:simulation_results}
In this section, the proposed joint scheduling strategy is evaluated with respect to its ability to reduce end-to-end latency by overlapping sensing and DNN execution. The performance of the proposed joint scheduling algorithm is further benchmarked against that of a conventional decoupled baseline, in which sensing and inference scheduling are performed separately.
\subsection{Simulation Setup}
In this subsection, the main simulation parameters are specified to characterize the considered multi-band radar sensing and DNN inference system. Unless otherwise specified, the same parameter setting is adopted in the illustrative scheduling example, and selected parameters will be varied later for quantitative performance comparison. As listed in Table~\ref{tab:simulation_parameters}, the considered system consists of \(K=6\) sensing bands and \(C=4\) accelerator cores. The corresponding multi-branch inference DAG is shown in Fig.~\ref{Fig2}. Each sensing band activates one branch of the DNN, and the branch outputs are merged through two alignment nodes, followed by a fusion head composed of \texttt{Fusion}, \texttt{Classifier}, and \texttt{Output} nodes. The numbers of nodes in the six branches are set to \([5,\,6,\,7,\,6,\,8,\,6]\), thereby introducing heterogeneous branch depths and different sensitivities to release times. The six branches are further divided into two alignment groups, where Slice~1--3 are connected to Align~1 and Slice~4--6 are connected to Align~2.

The sensing-information thresholds of the six bands are set to \([0.2,\,1.5,\,5,\,2,\,7,\,10]\) in normalized information units. The sensing process evolves over \(T_{\max}=2000\) time slots, with at most one radar band activated in each slot. The instantaneous sensing quality of each band is modeled by a time-varying SINR process, where the SINR values are independently generated within the range \([5,20]\) dB. A band is feasible for sensing only when its instantaneous SINR is no smaller than the threshold \(6\) dB. For each feasible sensing action, the information gain accumulated in that slot is computed as \(g_k(\gamma_k(t))=B_k\log_2\!\bigl(1+\gamma_k(t)\bigr)\), where \(B_k=180\) kHz denotes the effective sensing bandwidth of band \(k\). To avoid overly optimistic values, the effective spectral efficiency is further capped at \(8\) bit/s/Hz.

For the computation model, the execution time of each DAG node is randomly generated in \([1,11]\) ms. The on-chip read/write latency is generated in \([0.1,0.6]\) ms, while the off-chip read/write latency is generated in \([2,8]\) ms. These settings capture the tradeoff between data locality and execution parallelism: assigning dependent nodes to the same core reduces data-movement overhead, whereas distributing them across different cores may increase concurrency at the cost of off-chip memory access. Unless otherwise specified, the same random realization is adopted for both the proposed method and the decoupled baseline to ensure a fair comparison.
\begin{table}[t]
\caption{Simulation Parameters}
\label{tab:simulation_parameters}
\scriptsize
\centering
\begin{tabular}{ll}
\toprule
\textbf{Parameter} & \textbf{Value} \\
\midrule
Number of sensing bands \(K\) & \(6\) \\
Number of accelerator cores \(C\) & \(4\) \\
Number of nodes in each branch & \([5,\,6,\,7,\,6,\,8,\,6]\) \\
Band sensing-information thresholds \(\eta_k\) & \([0.2,\,1.5,\,5,\,2,\,7,\,10]\) \\
Alignment group assignment & \([1,\,1,\,1,\,2,\,2,\,2]\) \\
Fusion head & Enabled \\
Maximum number of time slots \(T_{\max}\) & \(2000\) \\
Maximum number of activated bands per slot & \(1\) \\
SINR threshold & \(6\) dB \\
Instantaneous SINR model & Uniformly generated in \([5,20]\) dB \\
Effective sensing bandwidth \(B_k\) & \(180\) kHz \\
Per-slot sensing information gain & \(g_k(\gamma_k(t))=B_k\log_2(1+\gamma_k(t))\) \\
Maximum effective spectral efficiency & \(8\) bit/s/Hz \\
Node computation time & Uniformly generated in \([1,11]\) ms \\
On-chip read latency & Uniformly generated in \([0.1,0.6]\) ms \\
On-chip write latency & Uniformly generated in \([0.1,0.6]\) ms \\
Off-chip read latency & Uniformly generated in \([2,8]\) ms \\
Off-chip write latency & Uniformly generated in \([2,8]\) ms \\
\bottomrule
\end{tabular}
\end{table}

\subsection{Illustrative Scheduling Example}
\begin{figure}
\centering
\includegraphics[width= 3.5 in]{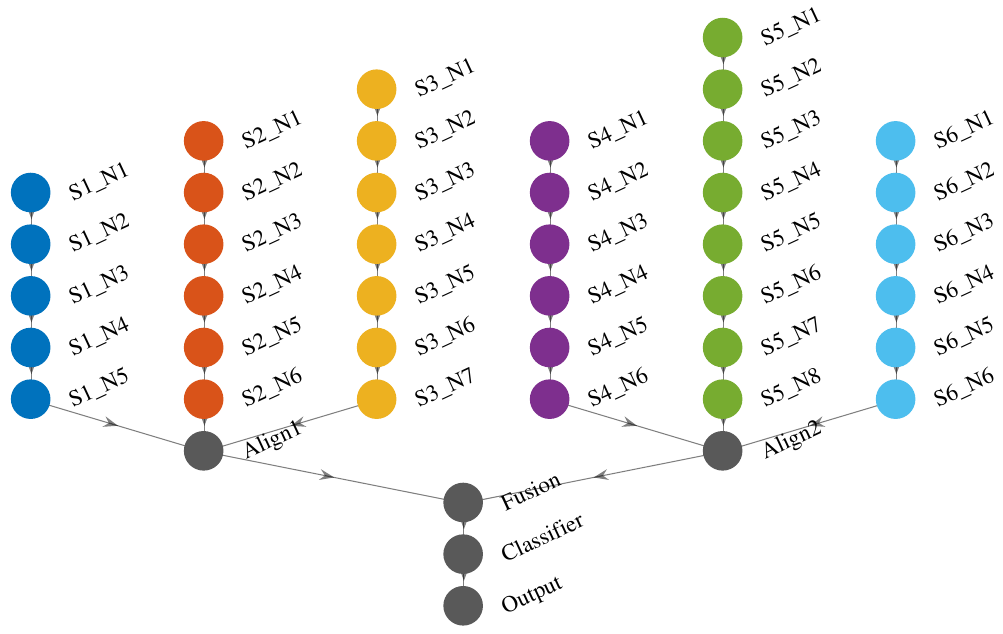}
\caption{A random DAG for the simulation.}
\label{Fig2}
\end{figure}
\begin{figure*}[t] 
\centering
\hspace*{-0.6in}
\includegraphics[width= 8.2 in]{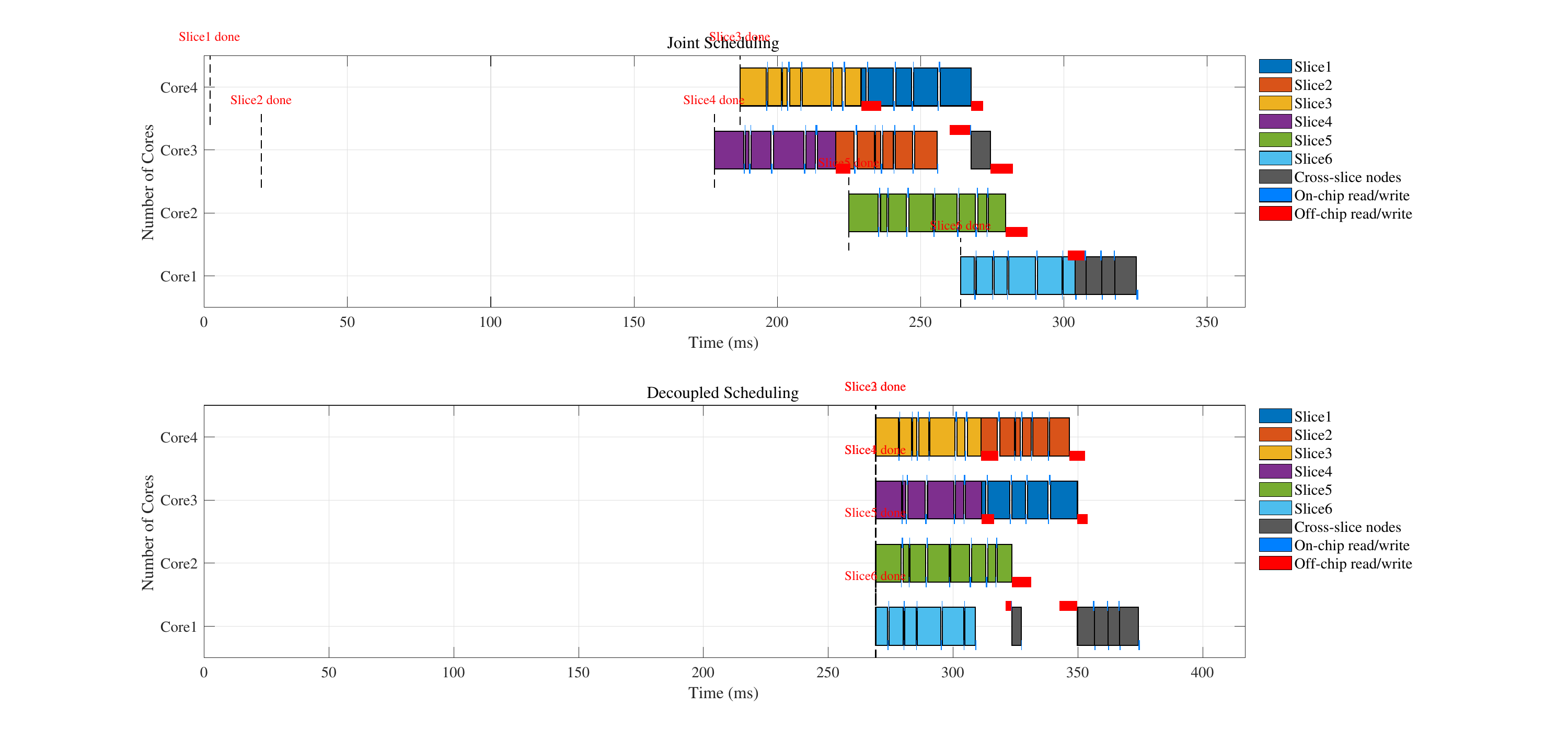}
\caption{Comparison of end-to-end execution timelines under the proposed joint scheduling method and the decoupled baseline.}
\label{Fig3}
\end{figure*}
Fig.~\ref{Fig2} shows the multi-branch DAG used in the simulation. Each colored chain corresponds to one sensing band and its associated inference branch, while the gray nodes denote the cross-branch modules, including the alignment nodes and the fusion head. This structure is representative of a multi-band inference pipeline, in which branch-specific computation can start as soon as the corresponding radar band completes sensing, whereas the final fusion stage must wait for the outputs of all required branches.

Fig.~\ref{Fig3} compares the execution timelines of the proposed joint scheduling method \textit{Joint Scheduling} and the decoupled baseline \textit{Decoupled Scheduling}. Different colors represent different branches, gray blocks denote cross-branch nodes, blue marks indicate on-chip read/write operations, and red marks indicate off-chip read/write operations. The dashed vertical lines indicate the release times of the branch-entry nodes, i.e., the instants at which the corresponding radar bands complete sensing. Fig.~\ref{Fig3} reveals three main observations. First, under the proposed method, branch execution starts immediately after the corresponding sensing band is completed, which enables effective overlap between sensing and DNN execution. By contrast, in the decoupled baseline, all branch-entry nodes wait until the last sensing band completes, so the inference stage starts later and the opportunity for pipelined execution is lost. Second, the proposed scheduler yields a more favorable release pattern for multi-core execution, since each sensing decision is made according to its impact on the final DAG makespan. As a result, the release of branch computations is better aligned with the DAG structure and core availability. Third, the overall completion time is visibly shortened under the proposed method.
\subsection{Impact of System-Scaling Factors}
\begin{figure*}[t]
\centering
\includegraphics[width=7.0in,trim=0cm 0cm 0cm 0cm,clip]{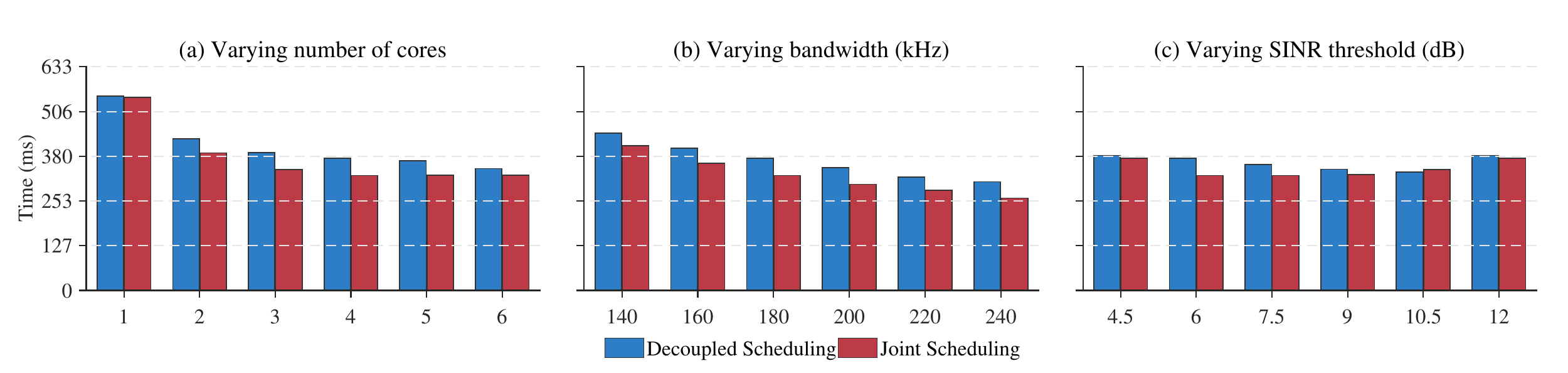}
\caption{The effects of three system-scaling dimensions—(a) the number of accelerator cores, (b) the bandwidth (kHz), and (c) the SINR threshold (dB).}
\label{Fig6}
\end{figure*}

Fig.~\ref{Fig6} illustrates the impact of three scaling factors, namely, the number of accelerator cores, the effective sensing bandwidth, and the SINR threshold, on the end-to-end latency. As the number of cores increases, the latency of both schemes decreases because more branch jobs can be processed in parallel; however, the gain gradually diminishes at large core counts, where the bottleneck shifts from computation parallelism to sensing release times and cross-branch synchronization. Increasing the effective sensing bandwidth increases the per-slot sensing information gain and correspondingly advances branch release times. Under this setting, the proposed \textit{Joint Scheduling} scheme consistently outperforms the \textit{Decoupled Scheduling} baseline, since it schedules band activations in a release-aware manner rather than treating bands independently of the downstream DAG structure.

\begin{table*}[t]
\centering
\caption{Total latency $T_{\mathrm{total}}$ (ms) under different slice-number, data-size, and core-number settings.}
\label{tab:latency_data_core}

\setlength{\tabcolsep}{2.6pt}
\renewcommand{\arraystretch}{1.12}
\scriptsize

\begin{tabular}{@{}>{\raggedright\arraybackslash}p{0.22\linewidth} ccc ccc ccc ccc@{}}
\toprule
\multicolumn{1}{c}{\multirow{2}{*}{Varying slice number and data sizes}}
& \multicolumn{3}{c}{\shortstack{C = 2 }}
& \multicolumn{3}{c}{\shortstack{C = 4 }}
& \multicolumn{3}{c}{\shortstack{C = 6 }}
& \multicolumn{3}{c}{\shortstack{C = 8 }}\\
\cmidrule(lr){2-4}\cmidrule(lr){5-7}\cmidrule(lr){8-10}\cmidrule(lr){11-13}
& \textbf{Joint} & \textbf{Decoupled} & \textbf{Gain}
& \textbf{Joint} & \textbf{Decoupled} & \textbf{Gain}
& \textbf{Joint} & \textbf{Decoupled} & \textbf{Gain}
& \textbf{Joint} & \textbf{Decoupled} & \textbf{Gain} \\
\midrule
\([0.2,\,1.5,\,5,\,2,\,7,\,10] \times 5\) kB
& 1415.27 & 1444.55 & +2.03\%
& 1413.27 & 1390.16 & -1.66\%
& 1414.27 & 1360.96 & -3.92\%
& 1414.27 & 1360.96 & -3.92\% \\

\([0.2,\,1.5,\,5,\,2,\,7,\,10]\) kB
& 388.94 & 428.55 & +9.24\%
& 325.27 & 374.16 & +13.07\%
& 326.27 & 344.96 & +5.42\%
& 326.27 & 344.96 & +5.42\% \\

\([0.2,\,1.5,\,5,\,2,\,7,\,10]/2\) kB
& 287.76 & 301.55 & +4.57\%
& 202.27 & 247.16 & +18.16\%
& 202.27 & 217.96 & +7.20\%
& 202.27 & 217.96 & +7.20\% \\

\([4,\,4,\,4,\,4,\,4,\,4]\) kB
& 401.55 & 407.55 & +1.47\%
& 350.98 & 353.16 & +0.62\%
& 322.96 & 323.96 & +0.31\%
& 322.96 & 323.96 & +0.31\% \\

\([0.2,\,\cancel{1.5},\,5,\,\cancel{2},\,7,\,\cancel{10}]\) kB
& 205.99 & 227.83 & +9.59\%
& 205.99 & 193.99 & -6.19\%
& 205.99 & 193.99 & -6.19\%
& 205.99 & 193.99 & -6.19\% \\
\bottomrule
\end{tabular}
\end{table*}

\subsection{Sensitivity to Slice-Data Heterogeneity}
Table~\ref{tab:latency_data_core} compares the total latency under different slice-data configurations and numbers of accelerator cores. Several observations can be made. For the heterogeneous and moderate-data setting (e.g., \([0.2,\,1.5,\,5,\,2,\,7,\,10]\)~kB), the proposed joint scheduling method achieves the most significant improvement, with a latency reduction of up to approximately $13.07\%$ at $C=4$. This is because the asynchronous branch release pattern in this regime is sufficiently staggered to create opportunities for overlap-aware scheduling, while still maintaining enough interaction among branches for joint optimization to be effective. By contrast, when all slices carry more data (the $\times$5 case), sensing completes slowly for most branches, and the performance gap narrows; at higher core counts, the decoupled scheme may even slightly outperform the greedy joint policy, indicating that the system bottleneck shifts from sensing--computation coordination to sensing side. When the data sizes are decreased (the /2 case), sensing becomes the dominant contributor to the end-to-end latency, and the proposed release-aware design yields more pronounced gains, reaching about $18.16\%$ at $C=4$. For the near-uniform configuration $[4,\,4,\,4,\,4,\,4,\,4]$~kB, the improvement remains positive but relatively small, suggesting that the advantage of joint scheduling becomes more evident when inter-slice heterogeneity is stronger. The last row further reports the simulation results for a reduced-slice configuration, where the original six-slice setting is simplified to three active slices, i.e., $[0.2,\,\cancel{1.5},\,\cancel{5},\,2,\, 7,\,\cancel{10}]$~kB. In this case, the overall latency is substantially reduced because fewer branches participate in both sensing and computation, thereby alleviating contention in sensing scheduling and accelerator execution. However, the benefit of the proposed joint method becomes less consistent: although it still achieves a gain of about $9.59\%$ at $C=2$, the decoupled baseline slightly outperforms it when $C\geq 4$. This result indicates that, as the number of slices decreases, the importance of cross-branch coordination is weakened, and the advantage of explicitly coupling sensing completion with DAG execution correspondingly diminishes. Overall, these results confirm that joint scheduling is particularly beneficial in heterogeneous, sensing-constrained, and sufficiently coupled multi-branch settings.

\begin{table*}[t]
\centering
\caption{The total latency $T_{\mathrm{total}}$ (ms) under different slice node-number settings.}
\label{tab:latency_nodes_core}

\setlength{\tabcolsep}{2.6pt}
\renewcommand{\arraystretch}{1.12}
\scriptsize

\begin{tabular}{@{}>{\raggedright\arraybackslash}p{0.22\linewidth} ccc ccc ccc ccc@{}}
\toprule
\multicolumn{1}{c}{\multirow{2}{*}{Varying slice node number}}
& \multicolumn{3}{c}{\shortstack{C = 2 }}
& \multicolumn{3}{c}{\shortstack{C = 4 }}
& \multicolumn{3}{c}{\shortstack{C = 6 }}
& \multicolumn{3}{c}{\shortstack{C = 8 }}\\
\cmidrule(lr){2-4}\cmidrule(lr){5-7}\cmidrule(lr){8-10}\cmidrule(lr){11-13}
& \textbf{Joint} & \textbf{Decoupled} & \textbf{Gain}
& \textbf{Joint} & \textbf{Decoupled} & \textbf{Gain}
& \textbf{Joint} & \textbf{Decoupled} & \textbf{Gain}
& \textbf{Joint} & \textbf{Decoupled} & \textbf{Gain} \\
\midrule
\([5,\,5,\,5,\,5,\,5,\,5]\)  
& 359.84  & 397.27 & +2.03\%
& 320.19 & 358.24 & -1.66\%
& 320.19  & 330.59 & -3.92\%
& 320.19  & 330.59 & -3.92\% \\

\([8,\,8,\,8,\,8,\,8,\,8]\)  
& 412.66 & 503.66 & +9.24\%
& 345.44 & 396.54 & +13.07\%
& 345.44 & 350.44 & +5.42\%
& 345.44 & 350.44 & +5.42\% \\

\([2,\,8,\,4,\,8,\,6,\,5]\)  
& 381.64  & 415.79 & +4.57\%
& 321.72 & 375.89 & +18.16\%
& 321.72 & 352.55 & +7.20\%
& 321.72 & 352.55 & +7.20\% \\

\([2,\,2,\,2,\,8,\,8,\,8]\)  
& 350.85 & 390.39 & +1.47\%
& 342.35 & 346.88 & +0.62\%
& 342.35 & 346.88 & +0.31\%
& 342.35 & 346.88 & +0.31\% \\

\([2,\,8,\,2,\,8,\,2,\,8]\)  
& 359.16 & 409.90 & +9.59\%
& 340.35 & 357.10 & -6.19\%
& 340.35 & 345.35 & -6.19\%
& 340.35 & 345.35 & -6.19\% \\
\bottomrule
\end{tabular}
\end{table*}

\subsection{Sensitivity to DAG Branch Depth}
Table~\ref{tab:latency_nodes_core} evaluates the total latency under different branch-depth configurations and numbers of accelerator cores. It can be observed that the proposed joint scheduling method generally achieves larger gains when the branch structures are more heterogeneous. For instance, in the mixed-depth configuration $[2,\,8,\,4,\,8,\,6,\,5]$, the latency reduction reaches approximately $18.16\%$ at $C=4$, indicating that the proposed policy can effectively exploit the staggered release patterns induced by different computational depths across branches. By contrast, for more regular structures such as $[2,\,2,\,2,\,8,\,8,\,8]$, the gain remains positive but relatively limited, since the execution patterns and synchronization points become less diverse, thereby reducing the opportunity for sensing--computation co-optimization. Another notable observation is that increasing the number of cores does not always lead to proportional latency reduction. In several cases, the performance improvement gradually saturates once $C$ exceeds 4--6, because the bottleneck shifts from per-core computation contention to branch release ordering and the critical path associated with alignment and fusion nodes. This trend is further reflected in scenarios where the decoupled baseline slightly outperforms the greedy joint policy at large $C$, suggesting that, under abundant computational resources, the residual performance gap is governed more by subtle global ordering effects than by raw parallel processing capacity. In summary, across both data-heterogeneity and node-heterogeneity settings, the proposed method reduces end-to-end latency in most practical operating points, and the gain is especially significant when sensing and computation are strongly coupled.

\section{Conclusions and Future Work}\label{sec:conclusions}
This paper studied latency-aware joint scheduling for multi-band radar sensing and DNN inference, and proposed a unified framework that captures the coupling between sensing-induced release times and multi-core DAG execution. The results show that end-to-end latency is mainly governed by the interaction between sensing completion and downstream computation, rather than by optimizing sensing or inference separately. Based on this insight, a release-aware greedy scheduler was developed to trigger branch execution earlier and increase temporal overlap between sensing and inference. Simulations showed that the proposed method can reduce latency compared with a decoupled baseline, achieving up to about 18\% improvement in heterogeneous and sensing-constrained settings. However, the gain becomes limited when sensing dominates the bottleneck or computation resources are abundant, and the method does not consistently outperform the decoupled strategy in all cases. These results confirm the value of joint scheduling while also highlighting the need for more robust algorithms with stronger performance guarantees under diverse system configurations.

\ifCLASSOPTIONcaptionsoff
  \newpage
\fi


\begin{thebibliography}{1}
\bibitem{Smith2024DeepLearning}
J. W. Smith and M. Torlak, ``Deep-learning-based multiband signal fusion for 3-D SAR superresolution,'' \textit{IEEE Trans. Aerosp. Electron. Syst.}, vol. 60, no. 1, pp. 8--24, Feb. 2024.
\bibitem{Liang2024RadarSignal}
R. Liang and Y. Cen, ``Radar signal classification with multi-frequency multi-scale deformable convolutional networks and attention mechanisms,'' \textit{Remote Sens.}, vol. 16, no. 8, Art. no. 1431, Apr. 2024.
\bibitem{Martone2020PracticalAspects}
A. F. Martone \textit{et al.}, ``Practical aspects of cognitive radar,'' in \textit{Proc. IEEE Radar Conf. (RadarConf20)}, Florence, Italy, 2020, pp. 1--6.
\bibitem{Howard2023HybridCognition}
W. W. Howard and R. M. Buehrer, ``Hybrid cognition for target tracking in cognitive radar networks,'' \textit{IEEE Trans. Radar Syst.}, vol. 1, pp. 118--131, 2023.
\bibitem{Zhao2024DeepMultimodal}
F. Zhao, C. Zhang, B. Geng \textit{et al.}, ``Deep multimodal data fusion,'' \textit{ACM Comput. Surv.}, vol. 56, no. 9, pp. 1--38, 2024.
\bibitem{Gehrig2021CombiningEvents}
D. Gehrig, M. R\"uegg, M. Gehrig, J. Hidalgo-Carrio and D. Scaramuzza, ``Combining events and frames using recurrent asynchronous multimodal networks for monocular depth prediction,'' \textit{IEEE Robot. Autom. Lett.}, vol. 6, no. 2, pp. 2822--2829, Apr. 2021.
\bibitem{Shi2024StreamingFlow}
Y. Shi \textit{et al.}, ``StreamingFlow: Streaming occupancy forecasting with asynchronous multi-modal data streams via neural ordinary differential equation,'' in \textit{Proc. IEEE/CVF Conf. Comput. Vis. Pattern Recognit. (CVPR)}, Seattle, WA, USA, 2024, pp. 21433--21442.
\bibitem{You2023AcceleratingConvolutional}
Y. You, P. Liu, D.-Y. Hong, J.-J. Wu, and W.-C. Hsu, ``Accelerating convolutional neural networks via inter-operator scheduling,'' in \textit{Proc. IEEE 28th Int. Conf. Parallel Distrib. Syst. (ICPADS)}, Nanjing, China, 2023, pp. 916--923.
\bibitem{Ding2021IOSInterOperator}
Y. Ding, L. Zhu, Z. Jia, G. Pekhimenko, and S. Han, ``IOS: Inter-operator scheduler for CNN acceleration,'' in \textit{Proc. Mach. Learn. Syst. (MLSys)}, virtual, 2021.

\bibitem{Gao2026Optimizing}
Z. Nie, H. Wang, A. T. Chronopoulos, Z. Tang, K. Li, C. Liu, and Z. Xiao, ``Adaptive block-wise mapping with intra-block resource allocation for multi-DNN workloads on heterogeneous accelerator systems,'' \textit{IEEE Trans. Parallel Distrib. Syst.}, vol. 37, no. 4, pp. 1015–1031, Apr. 2026.
\bibitem{Xu2023CoScheduling}
H. Liang, X. Jiang, J. Liu, X. Luo, S. Liu, N. Guan, and W. Yi, ``New scheduling algorithm and analysis for partitioned periodic DAG tasks on multiprocessors,'' \textit{IEEE Trans. Parallel Distrib. Syst.}, vol. 36, no. 12, pp. 2621–2634, Dec. 2025.
\bibitem{Zhong2024MultipleInOne}
N. Zhong \textit{et al.}, ``Multiple-in-one photonic integrated transceiver for multi-chirp-rate \& multi-band ISAR system and coherent fusion processing,'' \textit{J. Lightw. Technol.}, vol. 42, no. 21, pp. 7434--7442, 2024.
\bibitem{Yang2026MultiChannelSuperResolution}
S. Yang \textit{et al.}, ``Multi-channel super-resolution reconstruction model based on dual-band weather radar fusion,'' \textit{Remote Sens.}, vol. 18, no. 7, Art. no. 991, 2026.
\bibitem{Zhang2024DualBandSAR}
H. Zhang \textit{et al.}, ``The dual-band SAR image fusion-based foliage-penetrating target detection method,'' \textit{IEEE Trans. Geosci. Remote Sens.}, vol. 62, Art. no. 5226513, 2024.
\bibitem{Jiang2022MultiSubbandRadar}
Y. Jiang, S. Tang, M. Lu, and L. Zhang, ``Multi-subband radar signal fusion processing based on deep neural network in low signal-to-noise ratio,'' \textit{Wireless Commun. Mobile Comput.}, vol. 2022, Art. no. 9518542, 2022.
\bibitem{Gong2024MultibandRadar}
C. Gong, W. Li, R. Lu, and R. Wang, ``Multiband radar signal fusion and extrapolation method based on transformer model,'' in \textit{Proc. IEEE Int. Conf. Signal, Inf. Data Process. (ICSIDP)}, 2024.
\bibitem{Zhu2022SARImage}
J. Zhu, J. Pan, W. Jiang, X. Yue, and P. Yin, ``SAR image fusion classification based on the decision-level combination of multi-band information,'' \textit{Remote Sens.}, vol. 14, no. 9, Art. no. 2243, 2022.
\bibitem{Han2025ViTKAN}
S. Han, D. Ren, F. Gao, J. Yang, and H. Ma, ``ViT--KAN synergistic fusion: A novel framework for parameter-efficient multi-band PolSAR land cover classification,'' \textit{Remote Sens.}, vol. 17, no. 8, Art. no. 1470, 2025.
\bibitem{Yang2023DualBandPolarimetric}
W. Yang, Q. Zhou, M. Yuan, Y. Li, Y. Wang, and L. Zhang, ``Dual-band polarimetric HRRP recognition via a brain-inspired multi-channel fusion feature extraction network,'' \textit{Front. Neurosci.}, vol. 17, Art. no. 1252179, 2023.
\bibitem{Yang2024DualBandHRRP}
W. Yang, Z. Qi, H. Wu, Y. Li, L. Zhang, and Y. Wang, ``Dual-band HRRP recognition via wavelet packet decomposition and redundancy reduction model,'' in \textit{Proc. IEEE Int. Conf. Signal, Inf. Data Process. (ICSIDP)}, 2024.
\bibitem{Wang2024DualBandHRRP}
W. Wang, Z. Qi, L. Wang, W. Yang, L. Zhang, and Y. Wang, ``Dual-band HRRP fusion recognition via wavelet decomposition embedded autoencoder,'' in \textit{Proc. IEEE Int. Conf. Signal, Inf. Data Process. (ICSIDP)}, 2024.
\bibitem{Zhou2024MissingModality}
Q. Zhou \textit{et al.}, ``Missing modality completion for multi-frequency radar HRRP recognition using GAN,'' in \textit{Proc. IEEE Int. Conf. Signal, Inf. Data Process. (ICSIDP)}, 2024.
\bibitem{Baek2020MultiNeural}
E. Baek, D. Kwon, and J. Kim, ``A multi-neural network acceleration architecture,'' in \textit{Proc. ACM/IEEE 47th Annu. Int. Symp. Comput. Archit. (ISCA)}, 2020, pp. 940--953.
\bibitem{Kao2022MagmaOptimization}
S.-C. Kao and T. Krishna, ``Magma: An optimization framework for mapping multiple DNNs on multiple accelerator cores,'' in \textit{Proc. IEEE Int. Symp. High-Performance Comput. Archit. (HPCA)}, 2022, pp. 814--830.
\bibitem{Zheng2023MemoryComputation}
S. Zheng, S. Chen, and Y. Liang, ``Memory and computation coordinated mapping of DNNs onto complex heterogeneous SoC,'' in \textit{Proc. ACM/IEEE 60th Design Autom. Conf. (DAC)}, 2023, pp. 1--6.
\bibitem{Kim2023MoCAMemoryCentric}
S. Kim, H. Genc, V. V. Nikiforov, K. Asanović, B. Nikolić, and Y. S. Shao, ``MoCA: Memory-centric, adaptive execution for multi-tenant deep neural networks,'' in \textit{Proc. IEEE Int. Symp. High-Performance Comput. Archit. (HPCA)}, 2023, pp. 828--841.
\bibitem{Kwon2021HeterogeneousDataflow}
H. Kwon, L. Lai, M. Pellauer, T. Krishna, Y.-H. Chen, and V. Chandra, ``Heterogeneous dataflow accelerators for multi-DNN workloads,'' in \textit{Proc. IEEE Int. Symp. High-Performance Comput. Archit. (HPCA)}, 2021, pp. 71--83.
\bibitem{Kim2023DREAMDynamic}
S. Kim, H. Kwon, J. Song, J. Jo, Y.-H. Chen, L. Lai, and V. Chandra, ``DREAM: A dynamic scheduler for dynamic real-time multi-model ML workloads,'' in \textit{Proc. 28th ACM Int. Conf. Archit. Support Program. Lang. Oper. Syst. (ASPLOS)}, vol. 4, 2023, pp. 73--86.
\bibitem{Fan2023SparseDySta}
H. Fan, S. I. Venieris, A. Kouris, and N. Lane, ``Sparse-DySta: Sparsity-aware dynamic and static scheduling for sparse multi-DNN workloads,'' in \textit{Proc. 56th Annu. IEEE/ACM Int. Symp. Microarchitecture (MICRO)}, 2023, pp. 353--366.
\bibitem{Zhou2025TaiChiEfficient}
X. Zhou \textit{et al.}, ``TaiChi: Efficient execution for multi-DNNs using graph-based scheduling,'' in \textit{Proc. Design, Autom. Test Europe Conf. Exhib. (DATE)}, Lyon, France, 2025, pp. 1--7.
\end{thebibliography}
\end{document}